\documentclass[conference]{IEEEtran}
\IEEEoverridecommandlockouts
\usepackage{cite}
\usepackage{amsmath,amssymb,amsfonts}
\usepackage{algorithmic}
\usepackage{graphicx}
\usepackage{textcomp}
\usepackage{xspace}
\usepackage{xcolor}
\def\BibTeX{{\rm B\kern-.05em{\sc i\kern-.025em b}\kern-.08em
    T\kern-.1667em\lower.7ex\hbox{E}\kern-.125emX}}
\usepackage{cancel}
\usepackage{soul}
\usepackage{enumitem}
\usepackage{url}
\usepackage[hidelinks]{hyperref}
\usepackage{multirow}
\usepackage{multicol}
\usepackage{threeparttable}
\usepackage{booktabs}
\usepackage{listings}
\usepackage{flushend}
\usepackage{svg}

\usepackage{tcolorbox}%
\definecolor{mycolor}{rgb}{0.122, 0.435, 0.698}
\definecolor{gray1}{gray}{0.3}

\definecolor{darkgreen}{rgb}{0.0, 0.5, 0.0}
\definecolor{darkred}{rgb}{0.82, 0.1, 0.26}

\newcommand{\result}[1]{%
\begin{tcolorbox}[colframe=mycolor,boxrule=0.5pt,arc=4pt,
      left=6pt,right=6pt,top=6pt,bottom=6pt,boxsep=0pt,width=\columnwidth]%
      {#1}
\end{tcolorbox}%
}

\newcommand{\sqlancer}{\textit{SQLancer}\xspace}
\newcommand{\sqlright}{\textit{SQLRight}\xspace}
\newcommand{\tool}{\textit{SQLancer+QPG}\xspace}
\newcommand{\qpg}{\textit{QPG}\xspace}
\newcommand{\toolmiddle}{\textit{$SQLancer+QPG_{r}$}\xspace}
\newcommand{\todo}[1]{}
\renewcommand{\todo}[1]{{\color{red}{#1}}}

\newcommand{\Autoref}[1]{%
  \begingroup%
  \def\chapterautorefname{Chapter}%
  \def\sectionautorefname{Section}%
  \def\subsectionautorefname{Subsection}%
  \autoref{#1}%
  \endgroup%
}

\newcommand{\numbugs}{53\xspace}
\newcommand{\qpcovfactor}{4.85--408.48$\times$\xspace}
\newcommand{\qpcovfactorsqlright}{7.46$\times$\xspace}
\newcommand{\qplenfactor}{1.59--3.79$\times$\xspace}
\newcommand{\qplenfactorsqlright}{2.16$\times$\xspace}

\begin{document}

\title{Testing Database Engines via Query Plan Guidance}

\author{\IEEEauthorblockN{Jinsheng Ba}
\IEEEauthorblockA{National University of Singapore}
\and
\IEEEauthorblockN{Manuel Rigger}
\IEEEauthorblockA{National University of Singapore}
}

\maketitle

\begin{abstract}
Database systems are widely used to store and query data.
Test oracles have been proposed to find logic bugs in such systems, that is, bugs that cause the database system to compute an incorrect result.
To realize a fully automated testing approach, such test oracles are paired with a test case generation technique; a test case refers to a database state and a query on which the test oracle can be applied.
In this work, we propose the concept of \emph{Query Plan Guidance (QPG)} for guiding automated testing towards ``interesting'' test cases.
SQL and other query languages are declarative.
Thus, to execute a query, the database system translates every operator in the source language to one of the potentially many so-called physical operators that can be executed; the tree of physical operators is referred to as the query plan.
Our intuition is that by steering testing towards exploring a variety of unique query plans, we also explore more interesting behaviors---some of which are potentially incorrect.
To this end, we propose a mutation technique that gradually applies promising mutations to the database state, causing the DBMS to create potentially unseen query plans for subsequent queries.
We applied our method to three mature, widely-used, and extensively-tested database systems---SQLite, TiDB, and CockroachDB---and found \numbugs unique, previously unknown bugs.
Our method exercises \qpcovfactor more unique query plans than a naive random generation method and \qpcovfactorsqlright more than a code coverage guidance method.
Since most database systems---including commercial ones---expose query plans to the user, we consider QPG a generally applicable, black-box approach and believe that the core idea could also be applied in other contexts (e.g., to measure the quality of a test suite).
\end{abstract}

\begin{IEEEkeywords}
automated testing, test case generation
\end{IEEEkeywords}

\section{Introduction}

Database Management Systems (DBMSs) are fundamental software systems used to store, retrieve, and run queries on data. They are used in almost every computing device~\cite{sqlitedeployed, tidbdeployed, cockdorachdbdeployed}, thus any bug has a potentially severe consequence.
Logic bugs, which refer to incorrect results returned by DBMSs, are a particularly challenging category of bugs to find as they silently compute an incorrect result---unlike, for example, crash bugs~\cite{yan2018snowtrail, zhong2020squirrel}, which cause the process to be terminated. Consider \autoref{lst:motivation}, where the \lstinline{SELECT} statement triggers a logic bug that causes the returned result to unexpectedly contain a record, while it should be empty. 
Finding such bugs requires a so-called test oracle, which validates the DBMS' result.
Recently, effective test oracles~\cite{Rigger2020TLP, Rigger2020NoREC, Rigger2020PQS} have been proposed that brought validating the results of such queries within reach. 

Besides a test oracle, automatically finding logic bugs requires a test case generation method. For finding logic bugs in DBMSs, a test case refers to a database state and a query on which the test oracle can be applied.
Test case generation techniques face two main challenges.
First, ``interesting'' test cases should be generated that stress various parts of the DBMS to increase the chance of finding bugs in them.
No clear definition or metric on what an interesting test case constitutes exists, as it is unknown in advance by which logic bugs a DBMS is affected.
Second, the test cases should be valid both syntactically and semantically while also corresponding to the structure imposed by the test oracle; for example, the NoREC oracle requires a query with a \lstinline{WHERE} clause, but no more complex clauses (\emph{e.g.}, \lstinline{HAVING} clauses)~\cite{Rigger2020NoREC} while also forbidding various functions and keywords from being used (\emph{e.g.}, aggregate functions). 

Both generation-based and mutation-based approaches have been proposed to be paired with the above test oracles~\cite{Rigger2020TLP, Rigger2020NoREC, Rigger2020PQS}.
SQLancer uses a generation-based approach in which test cases are generated adhering to the grammar of the respective SQL dialects as well as the constraints imposed by the test oracles.
Overall, this approach makes it likely to generate valid test cases; we observed that about 90\% of the queries generated by SQLancer for SQLite are valid.
However, the test case generation approach receives no guidance that could steer it towards producing interesting test cases.
Recently, SQLRight~\cite{liang:sqlright} was proposed to address this shortcoming.
SQLRight mutates test cases aiming to maximize the DBMS' covered code, thus building on the success of grey-box fuzzing~\cite{afl, libfuzzer}.
While SQLRight improved on SQLancer's test case generation in various metrics, code coverage alone was shown to be an imperfect proxy metric for DBMSs~\cite{apollo} and stateful systems in general~\cite{sgfuzz}, as it cannot precisely model the state of databases.
Despite using mutation operators that aim to maximize the validity of queries, SQLRight achieves a lower rate of valid queries of 40\%~\cite{liang:sqlright}.
Other test case generation approaches have been proposed that aim at finding crash bugs and thus disregard the test oracle's constraints, which is why we do not further consider them.
These include mutation-based approaches such as Squirrel~\cite{zhong2020squirrel} or DynSQL~\cite{jiangdynsql}, and generation-based ones such as SQLsmith~\cite{sqlsmith} or RAGS~\cite{slutz1998rags}.

In this paper, we propose \emph{Query Plan Guidance} (\qpg), a technique that utilizes query plans to guide the test-case generation process towards interesting test cases. A query plan is a tree of operations that describes how an SQL statement is executed by a DBMS.
It is readily provided by DBMSs---users can typically obtain a textual representation using an \lstinline{EXPLAIN} SQL statement---and is typically inspected by DBMS users for tuning the performance of queries. Our insight is that a query plan provides a compact and high-level summary of how a query is executed, therefore, covering more unique query plans increases the likelihood of finding logic bugs.
Consider \autoref{lst:motivation}, which illustrates two scenarios of executing test cases with SQLite. 
In the first scenario, the \lstinline{CREATE INDEX} statement highlighted in red is omitted, causing the \lstinline{SELECT} statement to return an empty result.
This result is expected, since  column \lstinline{c} in table \lstinline{t2} has no data and the join condition \lstinline{c=3} is false. 
In the second scenario, the \lstinline{CREATE INDEX} statement is executed, which causes SQLite to unexpectedly fetch the row $\{|1|2|\}$. An index is an auxiliary data structure used by queries~\cite{graefe1993query}, which should not have any semantic effect.
While in both scenarios, the same query is executed, the query plans shown below the test cases differ due to the two different database states.
The left query plan for the correct execution indicates that the records from table \lstinline{t2} are read sequentially (\lstinline{SCAN t2}).
In contrast, the right query plan indicates that the DBMS used the index to read the data (\lstinline{SCAN t2 USING COVERING INDEX i0}), which was incorrect. Besides indexes, various other factors can influence query plans (\emph{e.g.}, data characteristics).

\begin{figure}
\begin{lstlisting}[caption={A bug found by \qpg in SQLite due to an incorrect use of an index in combination with a JOIN. Given the same SELECT, the left query plan is produced if no index is present, while the right one uses the index.},captionpos=t, label=lst:motivation, escapeinside=@@]
CREATE TABLE t1(a INT, b INT);
INSERT INTO t1(a) VALUES(2);
CREATE TABLE t2(c INT);
CREATE TABLE t3(d INT);
INSERT INTO t3 VALUES(1);
@\color{red}{CREATE INDEX i0 ON t2(c) WHERE c=3;}@

SELECT * FROM t2 RIGHT JOIN t3 ON d<>0 LEFT JOIN t1 ON c=3 WHERE t1.a<>0; --  @\{\}@ @\oksymbol@ @\color{red}{\{|1|2|\}}@ @\bugsymbol@
--------------------------------------------------
QUERY PLAN 
WITHOUT INDEX i0:           WITH INDEX i0:
|--SCAN t2                  |--SCAN t2 USING
                                COVERING INDEX i0
|--SCAN t3                  |--SCAN t3
|--SCAN t1                  |--SCAN t1
`--RIGHT-JOIN t3            `--RIGHT-JOIN t3
   `--SCAN t3                  `--SCAN t3
\end{lstlisting}
\end{figure}

To generate valid queries that correspond to the oracles' constraints, we propose mutating the database state rather than the queries.
Specifically, we re-use the existing random grammar-based generation approach of SQLancer~\cite{Rigger2020TLP} to generate the queries.
However, we record all seen query plans for a given database state and mutate this state when no new query plans are observed, indicating that the current database state's potential for enabling unobserved query plans has been saturated. We modeled the decision-making process for selecting the most promising mutation---an SQL statement that modifies the database state---as a multi-armed bandit problem and assigned a high priority to the SQL statement that results in the most new query plans across all executions. The multi-armed bandit problem is a model in which a fixed limited set of resources have to be allocated between competing choices in a way that maximizes the expected gain~\cite{berry1985bandit}.
We implemented \qpg in \sqlancer and evaluated it on SQLite, TiDB, and CockroachDB. We found \numbugs unique, previously unknown bugs, all of which have been acknowledged by the developers. Of these, 35 have already been fixed. Three bugs in SQLite had been hidden for more than six years before we found them, despite the extensive existing testing efforts by the authors of SQLancer and SQLRight, demonstrating the practical need for a more efficient test case generation approach. To trigger many of the bugs, complex query plans are required, indicated by the average length of query plans being 2.47$\times$ longer than that of the previously found bugs. In terms of efficiency, our \qpg-based implementation covers \qpcovfactor more unique query plans than \sqlancer and \sqlright in 24 hours. 

Overall, we make the following contributions:
\begin{itemize}
    \item We studied the query plans of the queries in previously-found bugs to gauge the idea's potential;
    \item We propose \emph{Query Plan Guidance} as a general idea for utilizing query plans for testing;
    \item We propose a concrete testing approach that mutates database state rather than queries to be compatible with existing test oracles;
    \item We implemented and evaluated the approach, which has found \numbugs unique, previously unknown bugs in widely-used DBMSs.
\end{itemize}
\section{Background}

\emph{Database management systems.} Database Management Systems (DBMSs) serve as an interface between applications and back-end data, helping users to store, manipulate, and query data based on an abstract data model. The relational data model~\cite{e2009derivability} is the most common model that has been adopted by most modern DBMSs. In this paper, we focus on testing such relational DBMSs. 

\emph{Structured Query Language.} The most commonly used language for interacting with relational DBMSs is the Structured Query Language (SQL)~\cite{chamberlin1974sequel}, which has been standardized by ISO/IEC 9075. SQL consists of many types of statements\cite{sql92}, which can be classified into three main sub-languages: 
\begin{enumerate}
    \item Data Query Language (DQL), which provides a \lstinline{SELECT} statement to query data.
    \item Data Definition Language (DDL), which is used to create and modify the schemas of data objects, for example, \lstinline{CREATE}, \lstinline{DROP}, and \lstinline{ALERT}.
    \item Data Manipulation Language (DML), which is used to modify the contents of data objects, for example, \lstinline{INSERT} and \lstinline{UPDATE}.
\end{enumerate}
While DDL and DML statements can affect the database state, queries (\emph{i.e.}, DQL statements) typically cannot. Our test cases consist of DQL, DDL, and DML statements.

\emph{Query plans.}  A query plan is a tree of operations that describes how an SQL statement is executed by a specific DBMS. Although not specified by the standard, most mature relational DBMSs, including the 10 most popular relational DBMSs according to the DB-Engines ranking,\footnote{\url{https://db-engines.com/en/ranking/relational+dbms}} allow users to query a textual representation of a query plan by prefixing a query with \lstinline{EXPLAIN}.
DBMSs cannot always determine the most efficient query plan~\cite{leis2015good,gu2012testing}, requiring users to understand and optimize performance-critical queries (\emph{e.g.}, by providing hints to the DBMS) based on their query plans.
For a better debugging experience, exposed query plans may include additional information, such as the estimated cost or predicate expressions (\emph{e.g.}, used in \lstinline{WHERE} clauses).
Database literature distinguishes between logical and physical query plans~\cite{silberschatz2002database}, the latter which is typically exposed by the DBMSs.
While the logical query plan closely corresponds to the original declarative query, the physical query plan maps every logical operator to a so-called physical one that can be executed by the DBMS.
For example, to translate a read operation on a table, the DBMS might choose one of potentially multiple so-called physical access methods (\emph{e.g.}, a full table scan, or a partial scan with index).
Similarly, to join two tables, the DBMS might decide between multiple join algorithms (\emph{e.g.}, hash join or nested loop join)~\cite{silberschatz2002database}.
Various factors influence what query plan a DBMS derives for a given query, such as characteristics of the data stored in the database~\cite{poosala1997histogram}, the existence of auxiliary data structures (\emph{e.g.}, indexes)~\cite{graefe2011modern}, the tables as well as views present in the database, and configuration options.
In this work, we use query plans in a black-box way, that is, without regarding the semantics of operators to guide testing.

\emph{Logic bugs.} Logic bugs are bugs that cause a system to compute incorrect results. 
Recently, Rigger et al. proposed several oracles~\cite{Rigger2020NoREC,Rigger2020TLP,Rigger2020PQS} that have found hundreds of unique bugs in widely-used DBMSs. In this work, we used the two latest test oracles, which represent the state of the art. Ternary Logic Partitioning (TLP) expects a query and derives multiple more complex queries, each of which computes a partition of the result to then check whether their results are equivalent. For example, from \lstinline{SELECT * FROM t0} and a random predicate \lstinline{t0.c0>0}, TLP derives \lstinline{SELECT * FROM t0 WHERE (t0.c0>0)}, \lstinline{SELECT * FROM t0 WHERE NOT (t0.c0>0)}, and \lstinline{SELECT * FROM t0 WHERE (t0.c0>0) ISNULL}, whose combined records must be equivalent to the first query. Non-optimizing Reference Engine Construction (NoREC)~\cite{Rigger2020NoREC} checks for inconsistent results values of a predicate used in a query that the DBMS might optimize and one that is used in a query that is difficult to optimize. For example, for a predicate \lstinline{t0.c0>0}, NoREC compares the number of rows returned by a query \lstinline{SELECT * FROM t0 WHERE (t0.c0>0)} with how often \lstinline{TRUE} is contained in the result returned for \lstinline{SELECT (t0.c0>0) FROM t0}. Both oracles have constraints on the query formats. For example, NoREC requires a \lstinline{WHERE} clause, but forbids aggregate functions and other more complex clauses. In principle, our method can be paired with any oracle.

\section{Query Plan Study}\label{sec:queryplanstudy}

\begin{table}
    \caption{Subjects for the query plan study.}
    \label{tab:survey_subject}
    \centering
    \begin{tabular}{@{}llrl@{}}
        \toprule
       \textbf{DBMS} & \textbf{Version} & \textbf{LoC} & \textbf{EXPLAIN Statement}\\
       \midrule
        CockroachDB & 19.2.12 & 1.1M & EXPLAIN (OPT)...  \\
        DuckDB & 0.19 & 59K & EXPLAIN...  \\
        H2 & 2.0.202 & 0.3M & EXPLAIN...  \\
        MariaDB & 10.4.25 & 3.6M & EXPLAIN FORMAT='JSON'...  \\
        MySQL & 5.7.33 & 3.8M & EXPLAIN FORMAT='JSON'...  \\
        PostgreSQL & 11.16 & 1.4M & EXPLAIN (COSTS FALSE)... \\
        SQLite & 3.30.0 & 0.3M & EXPLAIN QUERY PLAN... \\
        TiDB & 3.0.12 & 0.8M & EXPLAIN... \\
        \bottomrule
    \end{tabular}
\end{table}

\begin{table}
    \caption{Query plans of the queries in previously-found bugs. Length indicates the average number of operations in a query plan.}
    \label{tab:survey_result}
    \centering
    \begin{tabular}{@{}lrrrr@{}}
        \toprule
        &  & \multicolumn{3}{c}{\textbf{Query Plans}}  \\
       \cmidrule{3-5}
       \textbf{DBMS} & \textbf{Bugs} & Sum & Unique & Length \\
       \midrule
        CockroachDB & 68 & 37 & 32 & 3.43 \\
        DuckDB      & 75 & 59 & 18 & 2.00 \\
        H2          & 19 & 10 & 7  & 3.70 \\
        MariaDB     & 7  & 5  & 5  & 1.00 \\
        MySQL       & 40 & 35 & 22 & 1.03 \\
        PostgreSQL  & 31 & 9  & 3  & 2.33 \\
        SQLite      & 193& 118& 62 & 2.14 \\
        TiDB        & 62 & 43 & 32 & 5.07 \\
        \bottomrule
        &&& \textbf{Avg:} & 2.59 \\
    \end{tabular}
\end{table}

To investigate the potential of using query plans as guidance, we studied the uniqueness and complexity of query plans of the queries in previously-found bugs. We hypothesized that we would see a wide variety of query plans, suggesting that a bug-finding technique optimized for exploring more unique query plans might be effective.

\emph{Subjects.} We chose the public bug reports from \sqlancer as our subjects. \sqlancer provides a public list\footnote{\url{https://github.com/sqlancer/bugs}} including all found bugs and corresponding test cases for 499 bug reports across 9 DBMSs. We excluded 4 bugs found in the DBMS TDEngine, as this DBMS does not expose query plans. The query plan of a given query can vary over versions; thus, to obtain accurate query plans, we chose the most relevant release versions when the corresponding bugs were found. The details of the chosen DBMSs are shown in \Autoref{tab:survey_subject}.

\emph{Obtaining query plans.} For all 495 bug-inducing test cases, we instrumented all queries (\emph{i.e.}, \lstinline{SELECT} statements) by using \emph{EXPLAIN} statements as listed in \Autoref{tab:survey_subject}.
Depending on the DBMS, query plans might include various additional auxiliary information. We identified three such types. One type is the estimated cost (\emph{e.g.}, in PostgreSQL), which differs for almost every query. The second type is expressions in \lstinline{WHERE} clauses, which are included in the query plan by some DBMSs (\emph{e.g.}, CockroachDB). The third type is random identifiers, which are used to distinguish operations in a query plan (\emph{e.g.}, MariaDB and MySQL). To exclude such auxiliary information, we accordingly adjusted the parameters of the \texttt{EXPLAIN} statements, as shown in \Autoref{tab:survey_subject}. Lastly, we removed the names of tables, views, and indexes of the obtained query plans to distinguish query plans based on their structure only. This was based on the intuition that two query plans with the same execution logic, but different table names, would be processed similarly by the DBMSs (\emph{e.g.}, \texttt{SCAN t1}, and \texttt{SCAN t2}).

\emph{Uniqueness analysis.} \Autoref{tab:survey_result} shows the query plan distribution. In total, we obtained 316 query plans, of which 57.28\% were unique. The number of query plans is lower than that of test cases because 1) not all test cases have queries and 2) some queries that previously exposed bugs were rejected by subsequent versions of the DBMSs. The minimal percentage of unique query plans is 30.51\% in DuckDB. The maximum one is 100.00\% in MariaDB, due to a low number of test cases. Overall, for the queries in previously-found bugs, the variety of different query plans indicates that covering a wider variety of query plans might increase the likelihood of discovering bugs.

\result{Query plans of the queries in previously-found bugs vary significantly, as 57.28\% of the query plans are unique.}

\emph{Complexity analysis.} We examined the complexity of the query plans of the queries in previously-found bugs. A query plan with many operations is due to a complex database state or query. For instance, in SQLite, a query plan that retrieves data from two tables requires at least three operations: \lstinline{SCAN table t0}, \lstinline{SCAN table t1}, and \lstinline{MERGE results}, which is more complex than \lstinline{SCAN table t0} alone.  As shown in the \emph{Length} column of \Autoref{tab:survey_result}, the average number of operations per query plan is 2.59, which illustrates that the majority of bug-related query plans are compact. We further found that the most frequent query plan across eight DBMSs is \lstinline{SCAN table t0}, which represents a sequential scan on a single table, without using an index. For example, in SQLite, 26 of 118 query plans consist of a single table scan. This demonstrates that the query plans for the previously-found bugs are simple. While this could indicate that, compact and simple query plans are sufficient to trigger these previously found bugs---as suggested by the small-scope hypothesis~\cite{andoni2003evaluating}---it could also be that existing approaches have focused their testing on simple queries and database states. We speculate that covering more complex query plans might increase the likelihood of discovering bugs.

\result{Query plans of the queries in previously-found bugs are compact and simple, as the average number of operations in a query plan is only 2.59.}
\section{Approach}\label{sec:approach}

\begin{figure}
    \centering
    \includegraphics[width=\columnwidth]{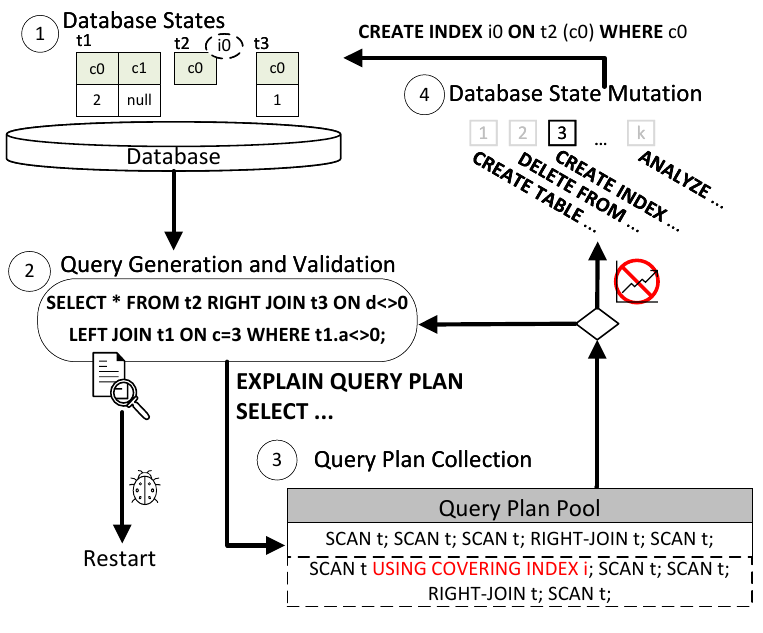}
    \vspace{-8mm}
    \caption{Overview of QPG. The dashed lines refer to the data affected by \textcircled{4} in the next iteration.}
    \label{fig:approach}
\end{figure}

To efficiently detect logic bugs in DBMSs, we propose to mutate databases with \emph{Query Plan Guidance (\qpg)} towards more unique and increasingly complex database states. Our insight is that the internal execution logic of the DBMS for a given query is reflected by its query plan and, therefore, covering more unique query plans increases the likelihood of finding logic bugs. Compared with naive random generation, our method gradually mutates database states enabling subsequent queries to cover more unique and complex query plans. We chose to mutate database states rather than queries, since test oracles have various constraints on queries, which are difficult to meet using mutational approaches~\cite{liang:sqlright}. Compared with other coverage-based grey-box testing tools for DBMSs, such as Squirrel\cite{zhong2020squirrel} and SQLRight~\cite{liang:sqlright}, we consider our method as black-box testing, as \qpg requires no access to the source code of the DBMS and uses information readily provided by mature DBMSs. Thus, the technique can also be applied to commercial closed-source DBMSs.

\emph{System overview.} \Autoref{fig:approach} shows an overview of our \qpg realization based on \Autoref{lst:motivation}. Given an initial database state at \textcircled{1}, \qpg generates a random SQL query at \textcircled{2} and executes it on the database to validate the query's result using the test oracle. If the oracle indicates a bug, \qpg outputs a bug report and restarts the testing process. Otherwise, it records the query plan and appends it to the query plan pool at \textcircled{3}. Typically, the execution continues at \textcircled{2} with the same database state. However, if no new unique query plan has been observed after a fixed number of iterations, \qpg mutates the database state at \textcircled{4} by applying a mutation operator to the current database state to create a new one, assuming that this new state will subsequently lead to new unique query plans being explored. 

\subsection{Database States. (\textcircled{1})}
The initial database state can be either randomly generated or manually given. In our implementation, we generate it by randomly executing DDL and DML statements. To avoid empty database states, we execute \lstinline{CREATE TABLE} statements first. For example, to create the initial database state in \Autoref{fig:approach}, we execute lines 1--5 in \Autoref{lst:motivation}. We do not directly manipulate database files, since they are highly structured~\cite{jeon2012recovery}, and any unexpected byte may incur an error that would impede the testing process.

\subsection{Query Generation and Validation. (\textcircled{2})}

\emph{Query generation.} We generate queries whose results we subsequently automatically validate to find bugs.
The generated queries must comply with two main constraints. First, queries must be semantically valid with respect to the database state. For example, they must reference only existing tables and views. Second, they must adhere to the constraints imposed by the test oracles. For example, the NoREC test oracle requires a \lstinline{WHERE} clause, but forbids other clauses (\emph{e.g.}, \lstinline{HAVING} or \lstinline{GROUP BY}). To address this, we adopt SQLancer's rule-based random generation approach that generates queries based on the SQL dialects' grammar adhering to the imposed constraints.
Many query generation approaches have been proposed~\cite{Bati2007, Bruno2006, apollo, Mishra2008, Poess2004, selinger1979, cockroachdb}, and our method can, in principle, be paired with any of these query generation methods.

\emph{Validation.} We use the state-of-the-art logic-bug oracles NoREC~\cite{Rigger2020NoREC} and TLP~\cite{Rigger2020TLP} to validate the queries' results. Both are metamorphic testing approaches~\cite{chen2018metamorphic} and, given a query, derive another query whose result set is used to validate the original query's result. In \Autoref{fig:approach}, given the three tables and the test oracle, we generate the query \lstinline{SELECT * FROM t2 RIGHT JOIN t3 ON d<>0 LEFT JOIN t1 ON c=3 WHERE t1.a<>0}. Since the test oracle indicates that the empty result returned is correct, execution continues at \textcircled{3}. If the test oracle indicates a bug, we output the bug report and restart the testing process.

\subsection{Query Plan Collection. (\textcircled{3})}
We collect query plans by instrumenting queries using \emph{EXPLAIN} statements, which is the same approach as presented in \Autoref{sec:queryplanstudy}. In \Autoref{fig:approach}, the statement to obtain the query plan is \lstinline{EXPLAIN QUERY PLAN SELECT * FROM t2 RIGHT JOIN t3 ON d<>0 LEFT JOIN t1 ON c=3 WHERE t1.a<>0}. We obtain the query plan (shown in the left part of lines 12--17 in \Autoref{lst:motivation}), and remove table and index names.

We insert query plans into the query plan pool in which we store unique query plans. The pool is implemented as a hash table in which the keys are query plans, and the values are the corresponding query strings. Given a query plan, we check whether the query plan exists in the pool, and insert it if not. 
In~\Autoref{fig:approach}, the pool is initially empty, so we insert the query plan (the first line at \textcircled{3}). 
If no new query plan is inserted into the pool for a fixed number of queries, we invoke \textcircled{4} aiming to cause the DBMS to explore more unique query plans. Otherwise, we continue to test the DBMS using the same database state at \textcircled{2}. 
A higher number indicates that we test the DBMS using more queries on a single database state, while a lower one means that we test the DBMS using more database states. The number is set to 1,000 by default, which we determined to work well empirically.

\subsection{Database State Mutation. (\textcircled{4})}
If no new query plan has been observed for a fixed number of queries, we invoke the database state mutation \textcircled{4}, which manipulates the database state, aiming to cause the DBMS to explore different query plans for the subsequent queries.

As mutation operators, we consider both the same DDL and DML statements used for generating the initial database state, such as \lstinline{CREATE TABLE}, \lstinline{CREATE INDEX}, and \lstinline{ANALYZE}.
A key challenge is to apply promising mutations that likely result in queries triggering new query plans. 
We model this task as the Multi-Armed Bandit (MAB) problem~\cite{gittins2011multi, berry1985bandit}, which is a popular and efficient method that has been used in various fuzzing works~\cite{yue2020ecofuzz, wang2021syzvegas, wu2022one, rebert2014optimizing}. In MAB, a fixed limited set of resources has to be allocated between competing choices to maximize the expected gain. In our scenario, given a limited computational resource, we choose the SQL statements (choices) to mutate database states to maximize the number of covered unique query plans (gain).

To maximize the expected gain, an automated agent attempts to acquire new knowledge (called ``exploration") and optimizes its decisions based on existing knowledge (called ``exploitation"). In our problem scenario, given the knowledge that the gains of only some mutation operators have been observed, we consider selecting the next mutation operator from either explored or unexplored mutation operators. Making the decision based on explored mutation operators (exploitation) tends to increase the gain, but may miss potentially higher gain from unexplored mutation operators. 
Many algorithms have been proposed to strike a balance between exploration and exploitation. We adopt the classic episode greedy algorithm~\cite{kuleshov2014algorithms}, which chooses the operator with the highest known gain with a certain probability and a random one otherwise.

Our algorithm works as follows. At $t$ times when database state mutation \textcircled{4} is invoked, we choose one mutation operator followed by~\Autoref{equ:greedy}. $k$ is the number of candidate mutation operators. $\hat{\mu}_{i(t)}$ is the known gain of the mutation operator $i$ at time $t$. $\epsilon$ is a fixed probability ranging from 0 to 1; its default value is 0.7,  which we determined to work well empirically. With $(1-\epsilon)$ probability, we choose the operator that has the maximum known gain and randomly choose one otherwise. 

\begin{equation}\label{equ:greedy}
\begin{split}
j(t) &= \begin{cases} 
            \arg \max _{i=1 \ldots k}\left(\hat{\mu}_{i(t)}\right) & (1 - \epsilon)\\
            random(k) & (\epsilon) \\
        \end{cases}
\end{split}
\end{equation}

\emph{Encoding known gain $\hat{\mu}_{i}$.} $\hat{\mu}_{i}$ is measured as weighted average gain---different from the standard algorithm, which uses an unweighted average---across all iterations where $i$ was chosen. A DBMS is a stateful system. The database state depends not only on the last applied mutation operator, but also on the previous database state. Applying the same mutation operator on changing database states creates different database states, so the gain of a mutation operator across iterations is not independent and identically distributed. For the same mutation operator, the gain in the last iteration is closer to the real gain on the last database state. To approximate the known gain, we use a weight average number in which the latter gain has a higher weight than the former gain. \Autoref{equ:avg} is our equation for updating $\hat{\mu}_{i}$ in each iteration. $Q$ is the gain for the last time $i$ was chosen. $w$ is the weight of $Q$, which is a constant ranging from 0 to 1; its default value is 0.25,  which we determined to work well empirically. Independent from the number of iterations, the prior gains only take up $(1-w)$ weight for $\hat{\mu}_{i}$. For example, given $w=0.1$, $\hat{\mu}_{i(999)}=0.1$, $Q=2$ for the $1,000_{th}$ iteration, the $\hat{\mu}_{i(1000)} = 0.1 + (2-0.1) * 0.1 = 0.29$, which is much higher than the unweighted average number $0.1 + (2-0.1)/1000 = 0.1019$ and closer to the $Q$. For efficiency, all parallel testing processes share the same $\hat{\mu}_{i}$.

\begin{equation}\label{equ:avg}
    \hat{\mu}_{i(t+1)} = \hat{\mu}_{i(t)} + (Q - \hat{\mu}_{i(t)}) * w
\end{equation}

\emph{Encoding instant gain $Q$.} $Q$ is measured by the proportion of queries that explore new query plans when they are executed on the latest database state. The queries include those in the query plan pool, and a set of newly generated queries based on the latest database state. The query plan pool includes all unique query plans and corresponding queries, which we re-execute to evaluate how many new query plans are explored for the same queries. To ensure that the queries in the query plan pool are always valid, we drop the invalid ones that are due to the changes of the database state. We observed that, in practice, this limits the pool to a reasonable size ($<8,000$ entries). However, for some mutation operators, such as \lstinline{CREATE TABLE}, none of these queries is related to the newly-created table, so no new query plan is observed. It would be unjust to judge its gain as zero, so we generate a set of new queries and examine how many new query plans are explored. For example, after applying the mutation operator $i$, $2/50$ queries in the query plan pool and $10/20$ queries in the set of newly generated queries explore unseen query plans, meaning that we compute the instant gain as $Q=2/50 + 10/20 = 0.54$.

\begin{figure}
    \centering
    \includegraphics[width=\columnwidth]{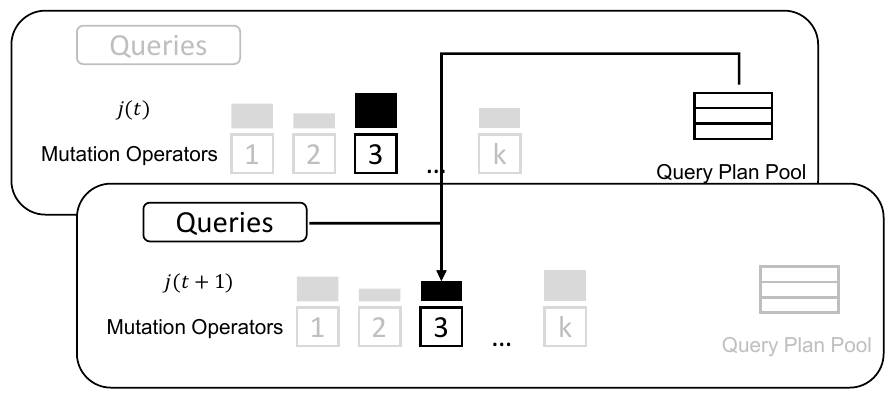}
    \caption{The workflow of measuring the known gain at \textcircled{4}.}
    \label{fig:mutationscheduler}
\end{figure}

\Autoref{fig:mutationscheduler} shows the workflow of measuring the known gain $\hat{\mu}_{i}$ at \textcircled{4}. If the mutation operator $3$ is chosen in iteration $t$ due to its highest $j(t)$, we update $\hat{\mu}_{3}$ in the next iteration $t+1$ with the queries that are generated after iteration $t$ and the queries of the query plan pool in iteration $t$. Following that, we calculate $j(t+1)$ and choose the mutation operator $k$.

In \Autoref{fig:approach}, we apply \lstinline{CREATE INDEX i0 ON t2 (c) WHERE c=3}, which creates an index \lstinline{i0} at \textcircled{1}. Suppose we generate the same query at \textcircled{2}, then we observe the new query plan shown on the right in lines 12--17 in \Autoref{lst:motivation} and insert it to the query plan pool. As a result, the bug is exposed at \textcircled{2}.

Lastly, we clear the database state after a fixed number of tested queries aiming to maximize the number of covered unique query plans. In general, by gradually mutating the same database state, we explore more unique and increasingly complex database states. However, the current database state may limit the possible state space to mutate into, which is why we clear the database state and restart the testing process after a fixed number of tested queries. The number is configurable and is set to a reasonable default value of 1,000,000, which we found to work well in our experiments (see \Autoref{subsec:Q4}).

\subsection{Implementation}
We implemented the described \qpg approach in \sqlancer\footnote{\url{https://github.com/sqlancer/sqlancer}} and subsequently refer to our prototype as \tool.
In addition, we updated SQLancer to support the latest version of SQLite which has three new features, namely \lstinline{RIGHT JOIN}, \lstinline{FULL OUTER JOIN}, and \lstinline{STRICT}.
We implemented our method in around 1,000 lines of Java code and adapted each DBMS-specific component in additional 100 lines of Java code, such as defining the specific statements for collecting query plans.
We designed our approach to be compatible with existing testing tools; thus, for the \emph{Database States}~\textcircled{1} and \emph{Query Generation and Validation}~\textcircled{2} steps, we reuse the implementation of \sqlancer.
We implemented the algorithm described in \emph{Database State Mutation}~\textcircled{4} as a standalone module that is reused across DBMSs.
We used DDL and DML statements supported by \sqlancer as mutation operators (23 mutations for SQLite, 13 mutations for TiDB, and 17 mutations for CockroachDB) which may contribute to covering more unique query plans, and the detailed list can be found in our artifact.
To avoid a large number of tables and indexes causing a low testing throughput, we restricted their maximum number to an arbitrary, but reasonable limit---a maximum of 10 tables and 20 indexes. 
\section{Evaluation}
To evaluate the effectiveness and efficiency of \qpg in finding bugs in DBMSs, we seek to answer the following questions based on our prototype \tool:
\begin{description}
    \item[\textbf{Q.1 New Bugs.}] Can \qpg help with finding new bugs? Are complex query plans required to find these bugs?
    \item[\textbf{Q.2 Covering unique query plans.}] Can \qpg cover more unique query plans than naive random generation and code-coverage guidance methods?
    \item[\textbf{Q.3 Bug Finding Efficiency.}] Can \qpg find bugs more efficiently than naive random generation and code-coverage guidance methods?
    \item[\textbf{Q.4 Sensitivity Analysis.}] What is the contribution of each component of \qpg? How does \qpg perform under different configurations?
\end{description}

\emph{Tested DBMSs.} We tested SQLite, TiDB, and CockroachDB. SQLite is the most popular embedded DBMS---embedded DBMSs are built together with and run in the same process as the application---and is used in every IOS and Android smartphone~\cite{sqlitedeployed}. TiDB and CockroachDB are popular enterprise-class DBMSs, and their open versions on Github are highly popular as they have been starred more than 31.9k and 25.2k times. They are widely used and have thus also been used in other DBMS testing works~\cite{liang:sqlright, Rigger2020PQS, Rigger2020TLP, zhong2020squirrel}. We did not consider other popular DBMSs due to various reasons. For example, for MySQL and closed-source DBMSs, bug fixes can be validated only after new releases; until then, it is difficult to identify new bugs, as already-known bugs might be repeatedly triggered. Furthermore, for some DBMSs, such as MySQL, many previously-reported bugs remain unfixed, impeding the testing process, which was also noted in prior work~\cite{Rigger2020TLP}. As a black-box method, \qpg supports any DBMS, regardless of what programming languages it is written in; SQLite is written in C, while TiDB and CockroachDB are written in Go.  
For Q1, Q2, and Q4, we used the latest available development versions (SQLite: 3.39.0, TiDB: 6.3.0, CockroachDB: 23.1). For Q3, to make a fair comparison, we chose the historical versions of DBMSs that all tools have tested and can find bugs in (SQLite: 3.36.0, TiDB: 4.0.15, and CockroachDB: 21.2.2).

\emph{Baselines.} We compared \tool with \sqlancer and \sqlright. While both of them have been designed to find logic bugs, their test case generation techniques differ. \sqlancer implements a naive random generation method. It is the baseline on which \tool is built. It has been starred more than 1,000 times on GitHub and is widely used by companies. \sqlright is the state-of-the-art tool and uses code-coverage guidance. By comparing with them, we gain insights into the benefits of \qpg against naive random generation and code-coverage-guided methods for finding logic bugs.

\emph{Experimental infrastructure.} We conducted all experiments on an Intel(R) Xeon(R) Gold 6230 processor that has 40 physical and 80 logical cores clocked at 2.10GHz. Our test machine uses Ubuntu 20.04 with 768 GB of RAM, and a maximum utilization of 40 cores. We repeated all experiments 10 times for statistically significant results.

\subsection*{Q.1 New Bugs}\label{sssec:Q1}

\newcommand{\numlogicbugs}{28\xspace}
\newcommand{\numerrorbugs}{25\xspace}

\begin{table}
    \centering
    \caption{The number of new bugs found by \tool.}
    \begin{tabular}{@{}lrrrr@{}}
        \toprule
       \textbf{DBMS} & \textbf{Crash} & \textbf{Error} & \textbf{Logic} & \textbf{All} \\
       \midrule
        SQLite & 0 & 5 & 23 & 28 \\
        TiDB & 2 & 4 & 3 & 9 \\
        CockroachDB & 3 & 11 & 2 & 16 \\
        \bottomrule
        \textbf{Sum:} & 5 & 20 & \numlogicbugs & \numbugs
    \end{tabular}
    \label{tab:bugs}
\end{table}

We ran \tool during approximately two months---during which we also implemented the approach---aiming to find bugs. To better demonstrate the underlying issue for each bug found, we minimized the test case both using C-Reduce\cite{regehr2012creduce} and manually. After reporting the bugs to the developers, we suspended the testing process until the bug was fixed to avoid duplicate reports whenever possible; when bugs were not fixed within a timespan of weeks, we reported multiple bugs that we suspected to be unique. The bugs in SQLite were usually fixed within 24 hours, while the bugs in TiDB and CockroachDB were usually fixed within several weeks. As a result, we focused on testing SQLite. We used NoREC~\cite{Rigger2020NoREC} and TLP~\cite{Rigger2020TLP}, which are the state-of-the-art oracles supported by both \sqlancer and \sqlright.

\emph{Bugs overview.} \Autoref{tab:bugs} shows the number of unique, previously unknown bugs found by \tool. We found \numbugs bugs in total, all of which have been confirmed. Of these, 35 have already been fixed. Although \sqlancer had been extensively applied to these DBMSs, we were still able to find these bugs with the help of \qpg. Of the \numbugs bugs, \numlogicbugs were logic bugs found by the test oracles TLP and NoREC, and \numerrorbugs bugs were associated with crashes or internal errors. This demonstrates that the complex database states generated by \qpg are beneficial not only to finding logic bugs, but also to other kinds of bugs. Although CockroachDB used the TLP oracle in their Continuous Integration (CI) process,\footnote{\url{https://github.com/cockroachdb/cockroach/commit/777382e6}} we still found 16 previously unknown bugs using \qpg. For the new features in SQLite, \qpg found 13 bugs in \lstinline{RIGHT JOIN}, 2 bugs in \lstinline{FULL JOIN}, and no bug in \lstinline{STRICT}. We give two examples of found bugs as follows. 

\begin{figure}
\begin{lstlisting}[caption={A bug in the RIGHT JOIN feature of SQLite.},captionpos=t, label=lst:case1, escapeinside=@@]
CREATE TABLE t1(a CHAR);
CREATE TABLE t2(b CHAR);
INSERT INTO t2 VALUES('x');
CREATE TABLE t3(c CHAR NOT NULL);
INSERT INTO t3 VALUES('y');
CREATE TABLE t4(d CHAR);

SELECT * FROM t4 LEFT OUTER JOIN t3 ON TRUE INNER JOIN t1 ON t3.c='' @\textbf{\color{red}{RIGHT OUTER JOIN}}@ t2 ON t3.c='' WHERE t3.c ISNULL;  -- {} @\bugsymbol@, {|||x} @\oksymbol@
\end{lstlisting}
\end{figure}

\emph{Example 1: a bug in the \lstinline{RIGHT JOIN} feature.} \Autoref{lst:case1} shows a test case exposing a logic bug that we found in SQLite. The \lstinline{SELECT} statement incorrectly returns an empty result, because of an incorrect optimization of \lstinline{ISNULL} when used with a \lstinline{RIGHT JOIN}. The query plan of the \lstinline{SELECT} statement is six operations long: scanning all tables once in four operations, and joining table \lstinline{t2} with another scan on \lstinline{t2} in two operations. 
The query plan is relatively long, because joining tables typically involves multiple operations. 13 bugs in SQLite were in the \lstinline{RIGHT JOIN} feature, in which \qpg generates more complex database states to find bugs.

\begin{figure}
\begin{lstlisting}[caption={A bug in \lstinline{json\_quote} function of SQLite.},captionpos=t, label=lst:case2, escapeinside=@@]
CREATE TABLE t1 (a CHAR);
CREATE VIEW v1(b) AS SELECT json(TRUE);
INSERT INTO t1 VALUES ('x');

SELECT * FROM v1, t1 WHERE NOT @\color{red}{json\_quote(b)}@; -- {} @\bugsymbol@, {1|x} @\oksymbol@
\end{lstlisting}
\end{figure}

\emph{Example 2: a bug in JSON feature.} \Autoref{lst:case2} is another logic bug that had existed in SQLite since July 23, 2016. The \lstinline{SELECT} statement incorrectly returns an empty result because of an incorrect optimization of the \lstinline{json_quote} function in the context of a \lstinline{VIEW}, which is necessary to find the bug. The bug cannot be found if the second line is replaced by \lstinline{CREATE TABLE v1(b) AS SELECT json(1)}. In SQLite, we found three bugs that had been hidden for more than six years, and \tool is the first tool to find them despite extensive efforts by the authors of \sqlancer and \sqlright.

\newcommand{\planlenbugs}{6.35\xspace}
\newcommand{\planuniqueper}{63.77\%\xspace}

\begin{table}
    \centering
    \caption{Query Plans of the queries in newly found bugs.}
    \begin{tabular}{@{}lrrr@{}}
        \toprule        

       \textbf{DBMS} & \textbf{All} & \textbf{Unique} & \textbf{Length} \\
       \midrule
        SQLite & 51 & 29  & 5.55 \\
        TiDB & 12 & 9 & 5.67 \\
        CockroachDB & 6 & 6 & 7.83 \\
        \bottomrule
        & & \textbf{Avg:} & \planlenbugs
    \end{tabular}
    \label{tab:bugs_queryplan}
\end{table}

\emph{The uniqueness and complexity of query plans.} To better understand how and whether \qpg enables exploring a variety of query plans, we analyzed the query plans of the queries in \Autoref{tab:bugs}. In total, we obtained 69 query plans, of which \planuniqueper are unique. This further demonstrates the diversity of query plans. On average, the length of query plans of queries was \planlenbugs. In comparison with \Autoref{tab:survey_result}, where the average number of operations in a query plan was 2.59, more complex query plans are required to expose these newly found bugs, and \qpg was successful in causing them to be generated. 

\result{With the help of \qpg, we found \numbugs unique, previously unknown bugs where the average length of query plans of queries is \planlenbugs.}

\subsection*{Q.2 Covering Unique Query Plans}\label{subsec:Q2}
We evaluated whether \tool can cover more unique query plans than \sqlancer and \sqlright in 24 hours. Our study in \Autoref{sec:queryplanstudy} shows that query plans in previously-found bugs are diverse, so covering more unique query plans likely increases the probability of finding bugs. We designed \tool to explore more unique and complex query plans than \sqlancer. We used the TLP oracle, which is the only test oracle that is supported by all DBMSs we considered.

\begin{figure}
    \centering
    \includegraphics[width=\columnwidth]{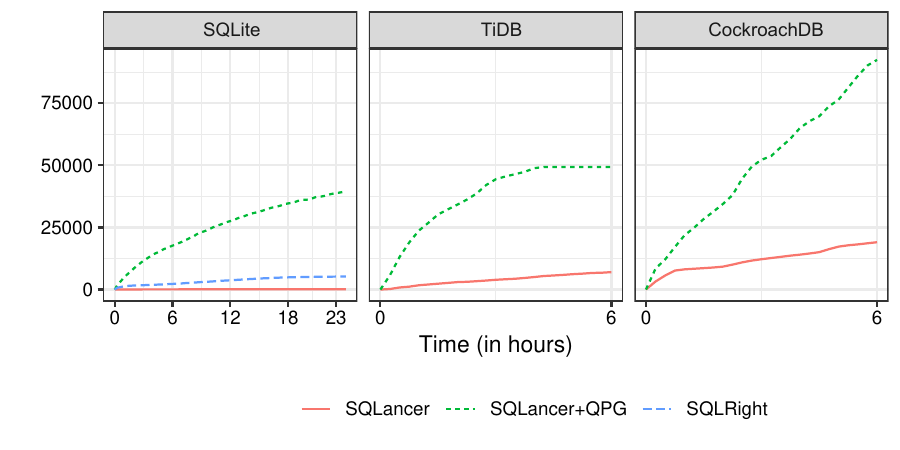}
    \vspace{-10mm}
    \caption{The average number of unique query plans across 10 runs in 24 hours. We could run TiDB and CockroachDB only for 6 hours due to crashes.}
    \label{fig:queryplancov}
\end{figure}

\begin{table}
    \centering
    \caption{The average and median number of query plan lengths across 10 runs in 24 hours. Only 6 hours are shown for TiDB and CockroachDB because of crashes.}
    \begin{tabular}{@{}lrr rr rr@{}}
        \toprule
        & \multicolumn{2}{c}{\textbf{\sqlancer}} & \multicolumn{2}{c}{\textbf{\sqlright}} & \multicolumn{2}{c}{\textbf{\tool}} \\
        \cmidrule(lr){2-3}
        \cmidrule(lr){4-5} 
        \cmidrule(lr){6-7}
        \textbf{DBMS} & \textbf{Avg} & \textbf{Median} & \textbf{Avg} & \textbf{Median} & \textbf{Avg} & \textbf{Median} \\
        \midrule
        SQLite & 2.95 & 2.00 & 2.17 & 1.00 & 4.69 & 4.00 \\
        TiDB & 3.97 & 2.00 & - & - & 15.04 & 8.20  \\
        CockroachDB & 4.55 & 4.00 & - & - & 8.87 & 6.90 \\
        \bottomrule
        \textbf{Avg:} & 3.82 & 2.67 & 2.17 & 1.00 & 9.53 & 6.37 \\
    \end{tabular}
    \label{tab:queryplan_len}
\end{table}

\emph{Measurements}. \Autoref{fig:queryplancov} shows the average number of unique query plans covered by all tools across 10 runs in 24 hours. We recorded the query plans every 15 minutes and removed the names of tables, views, and indexes as described in \Autoref{sec:queryplanstudy}. For TiDB and CockroachDB, we could run \tool at most for 6 hours, because \tool found several crash bugs that remained unfixed during our evaluation. We could run \sqlright only on SQLite, as \sqlright does not support TiDB and CockroachDB.
\Autoref{tab:queryplan_len} shows the average and median lengths of query plans of the queries executed across 10 runs in 24 hours. 

\emph{Results}. On both metrics, the number of unique query plans and their complexity, \tool clearly outperforms \sqlancer and \sqlright. 
\tool exercises \qpcovfactor more unique query plans than \sqlancer and \qpcovfactorsqlright more than \sqlright.  CockroachDB provides fine-grained query plans, which is why \tool most clearly outperformed \sqlancer on this DBMS. 
The growth rate of \tool in TiDB stagnates at around 5 hours due to a crash bug that terminated the TiDB server process. 
\Autoref{tab:queryplan_len} shows that the average length of query plans in \tool is \qplenfactor longer than for \sqlancer, and \qplenfactorsqlright longer than for \sqlright. 
To mitigate randomness, we measured the Vargha-Delaney\cite{vargha2000critique} ($\hat A_{12}$) and Wilcoxon rank-sum test\cite{mann1947test} ($U$) of \tool against \sqlancer. $\hat A_{12}$ measures the \emph{effect size} and gives the probability that random testing of \tool is better than random testing of \sqlancer (\emph{i.e.}, $\hat A_{12}>0.5$ means \tool is better). The Wilcoxon rank sum test $U$ is a non-parametric \emph{statistical hypothesis test} to assess whether the result differs across both tools. We reject the null hypothesis if $U<0.05$, that is, \tool outperforms \sqlancer with statistical significance. For both metrics, $\hat A_{12}=1$ and $U<0.05$ for \tool against \sqlancer on all DBMSs. The results show that our algorithm continuously generates significantly more unique and complex database states for testing.

\begin{table}
    \centering
    \caption{The line and branch coverage across 10 runs in 24 hours.}
    \begin{tabular}{@{}lrr rr rr@{}}
        \toprule
        & \multicolumn{2}{c}{\textbf{\sqlancer}} & \multicolumn{2}{c}{\textbf{\sqlright}} & \multicolumn{2}{c}{\textbf{\tool}} \\
        \cmidrule(lr){2-3}
        \cmidrule(lr){4-5}
        \cmidrule(lr){6-7}
        \textbf{DBMS} & \textbf{Line} & \textbf{Branch} & \textbf{Line} & \textbf{Branch} & \textbf{Line} & \textbf{Branch} \\
        \midrule
        SQLite & 30.3\% & 22.7\% & 48.1\% & 38.9\% & 32.6\% & 24.4\%\\
        \bottomrule
    \end{tabular}
    \label{tab:branch_cov}
\end{table}

\result{\qpg exercises  \qpcovfactor more unique query plans than a naive random generation method and \qpcovfactorsqlright more than a code-coverage guidance method.}

\emph{Code coverage.} While we were primarily interested in covering more unique query plans, code coverage is a common metric of interest that also gives some insights on how much of a system might be tested.
Thus, we evaluated the line and branch coverage of all three tools. 
Since TiDB and CockroachDB are written in Go, which is not supported by \sqlright, we measured code coverage only for SQLite. \Autoref{tab:branch_cov} shows the average percentage of line and branch coverage across 10 runs in 24 hours. Although \tool does not aim to maximize code coverage, \tool still outperforms \sqlancer on both line coverage and branch coverage because of more unique query plans covered. \sqlright clearly achieves the highest coverage. The reasons for this are that 1) \sqlright was designed to increase code coverage, 2) \sqlancer and \tool only generate SQL statements for the core logic of DBMS, while \sqlright produces all kinds of SQL statements by parsing the grammar files from DBMSs, and 3) \sqlright provides high-quality seeds that already cover 34.1\% line coverage and 26.4\% branch coverage, outperforming the other tools even without mutations. Since SQLite achieves 100\% branch coverage in their internal testing,\footnote{\url{https://www.sqlite.org/testing.html\#mcdc}} we believe that higher code coverage has a limited contribution for finding logic bugs.

\subsection*{Q.3 Bug Finding Efficiency}\label{subsec:Q3}
We evaluated whether \tool finds bugs faster than \sqlancer and \sqlright. To this end, we ran \tool, \sqlancer, and \sqlright for 24 hours with the TLP oracle. We used a best-effort method to distinguish unique bugs by checking whether 1) stack traces are the same (crash bugs); 2) error messages are the same (error bugs); 3) SQL clause structures are the same (logic bugs), such as two bugs' queries that only have \lstinline{RIGHT JOIN} and \lstinline{GROUP BY} clauses are deemed to be duplicate bugs.

\begin{table}
    \centering
    \caption{The number of all and unique bugs found across 10 runs.}
    \begin{tabular}{@{}lrr rr rr@{}}
        \toprule
        & \multicolumn{2}{c}{\textbf{\sqlancer}} & \multicolumn{2}{c}{\textbf{\sqlright}} & \multicolumn{2}{c}{\textbf{\tool}} \\
        \cmidrule(lr){2-3}
        \cmidrule(lr){4-5} 
        \cmidrule(lr){6-7}
        \textbf{DBMS} & \textbf{All} & \textbf{Unique} & \textbf{All} & \textbf{Unique} & \textbf{All} & \textbf{Unique} \\
        \midrule
        SQLite & 2 & 1 & 2 & 1 & 4 & 2 \\
        TiDB & 56 & 10 & - & - & 118 & 12 \\
        CockroachDB & 4 & 2 & - & - & 8 & 3 \\
        \bottomrule
        \textbf{Sum: } & 62 & 13 & 2 & 1 & 130 & 17
    \end{tabular}
    \label{tab:bugefficient}
\end{table}

\Autoref{tab:bugefficient} shows the sum of all bugs and only assumed-unique bugs found by each tool in 24 hours and 10 runs. Since crash bugs terminate the whole process, all experiments concluded in less than 24 hours until the first crash was observed (SQLite: 9 hours, TiDB: 1 hour, and CockroachDB: 16 hours). We did not restart the testing process as this would disadvantage \tool by making it lose the database states. Overall, \tool found 2$\times$ more bugs and 1.4$\times$ more unique bugs than \sqlancer; 65$\times$ more bugs and 17$\times$ more unique bugs than \sqlright. As duplicate bugs significantly slow down the testing process and hinder finding other bugs, the number of unique bugs is much smaller than the number of all bugs. In TiDB, we found several easy-to-reach bugs in \lstinline{JOIN}s, which do not require complex database states, so the number of all bugs is much higher than for the others. The results further show that bugs can be more efficiently found by exploring more unique query plans. 

\result{\qpg finds previous bugs 1.4$\times$ faster than a naive random generation method and 17$\times$ faster than a code-coverage guidance method.}

\subsection*{Q.4 Sensitivity Analysis}\label{subsec:Q4}
To evaluate the contribution of \tool's components, we performed a sensitivity analysis.

\begin{figure}
    \centering
    \includegraphics[width=\columnwidth]{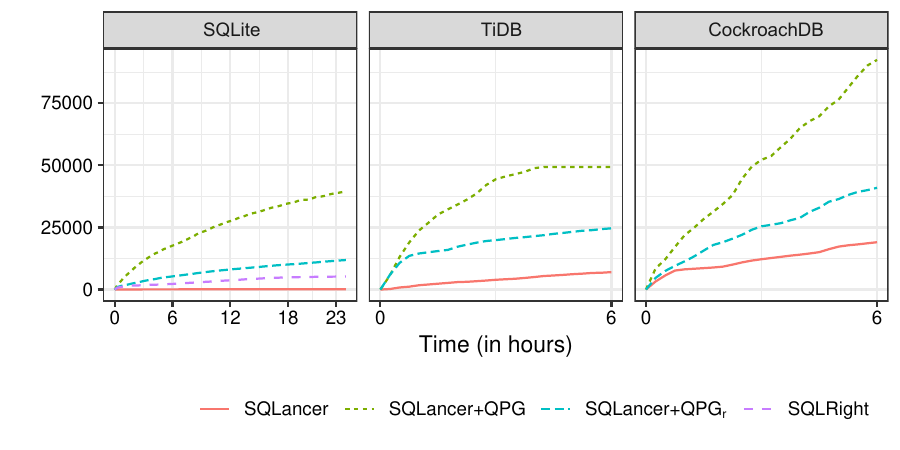}
    \vspace{-10mm}
    \caption{The average number of covered unique query plans to evaluate \textbf{the contributions of algorithm components} across 10 runs in 24 hours.}
    \label{fig:random}
\end{figure}

\begin{figure}
    \centering
    \includegraphics[width=0.8\columnwidth]{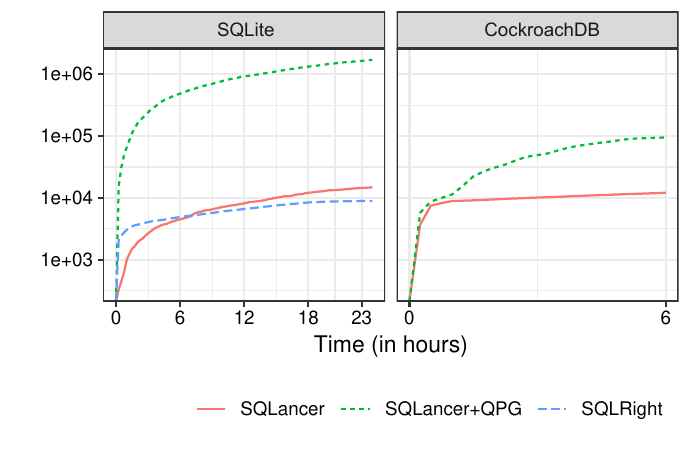}
    \vspace{-6mm}
    \caption{The average number of covered unique query plans by the \textbf{NoREC oracle} across 10 runs in 24 hours. The y axis uses a log scale.}
    \label{fig:norec}
\end{figure}

\emph{Contributions of algorithm components}. Our major contributions are \emph{query plan collection} \textcircled{3} and \emph{database state mutation} \textcircled{4} shown in  \Autoref{fig:approach}. To assess their contributions, we derived a new configuration \toolmiddle that enables only the query plan collection \textcircled{3}, and randomly applies mutations in \textcircled{4}. 
\Autoref{fig:random} shows the average number of covered unique query plans across 10 runs in 24 hours with the TLP oracle. \tool outperforms \toolmiddle, demonstrating the contribution of \textcircled{4}. \toolmiddle outperforms \sqlancer, demonstrating the contribution of \textcircled{3}. \tool has a higher growth rate than \toolmiddle, because \textcircled{4} gradually learns which mutation operators are promising. Due to the crash bugs, we ran TiDB and CockroachDB for only 6 hours.

\emph{Sensitivity of oracles.} We also evaluated \tool with NoREC, which is the second state-of-the-art oracle. 
\Autoref{fig:norec} shows the average number of covered unique query plans across 10 runs in 24 hours for NoREC oracle. \sqlancer lacks a NoREC oracle for TiDB, so we exclude it here. All tools have a higher number of covered unique query plans with the NoREC than with the TLP oracle, because of different constraints on queries from NoREC and TLP. Similar to TLP, \tool gains a significant advantage over \sqlancer and \sqlright with the NoREC oracle.

\begin{figure}
    \centering
    \includegraphics[width=\columnwidth]{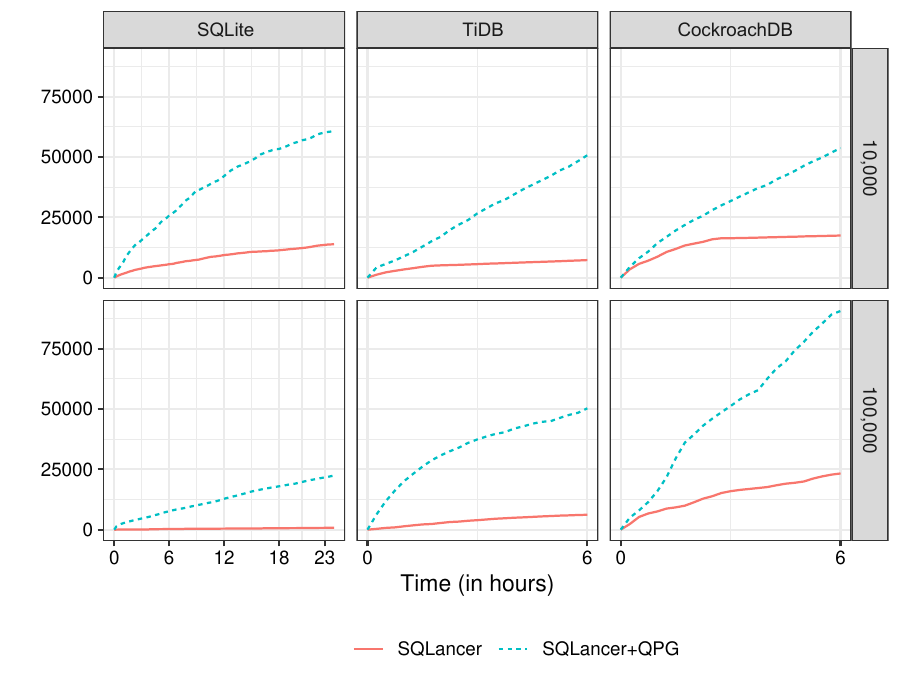}
    \vspace{-10mm}
    \caption{The average number of covered unique query plans by \textbf{varying the maximum number of queries per database state} across 10 runs in 24 hours.}
    \label{fig:num_queries}
\end{figure}

\emph{Sensitivity of maximum queries per database}. 
Both \tool and \sqlancer have a configuration to control the number of tested queries before clearing database states and starting a fresh testing instance.
The default value for both is 1,000,000. Often, starting a fresh testing instance at \textcircled{1} may result in a higher number of covered unique query plans. To evaluate whether \tool still performs well when more frequently resetting database states, we adjusted the number to 10,000 and 100,000, and evaluated the number of their covered unique query plans. \Autoref{fig:num_queries} shows the average number of covered unique query plans under the various maximum number of queries per database state. 
\tool gains a significant advantage over \sqlancer in all experiments. We clearly see that the rate of newly discovered query plans of \sqlancer stagnates over time, while \tool's rate continues to increase. Configuring the number is a trade-off since \tool creates more complex query plans with a higher number of maximum queries per database state and more unique query plans with a lower number.
A user can adjust the configuration option depending on the testing goals.

\begin{figure}
    \centering
    \includegraphics[width=\columnwidth]{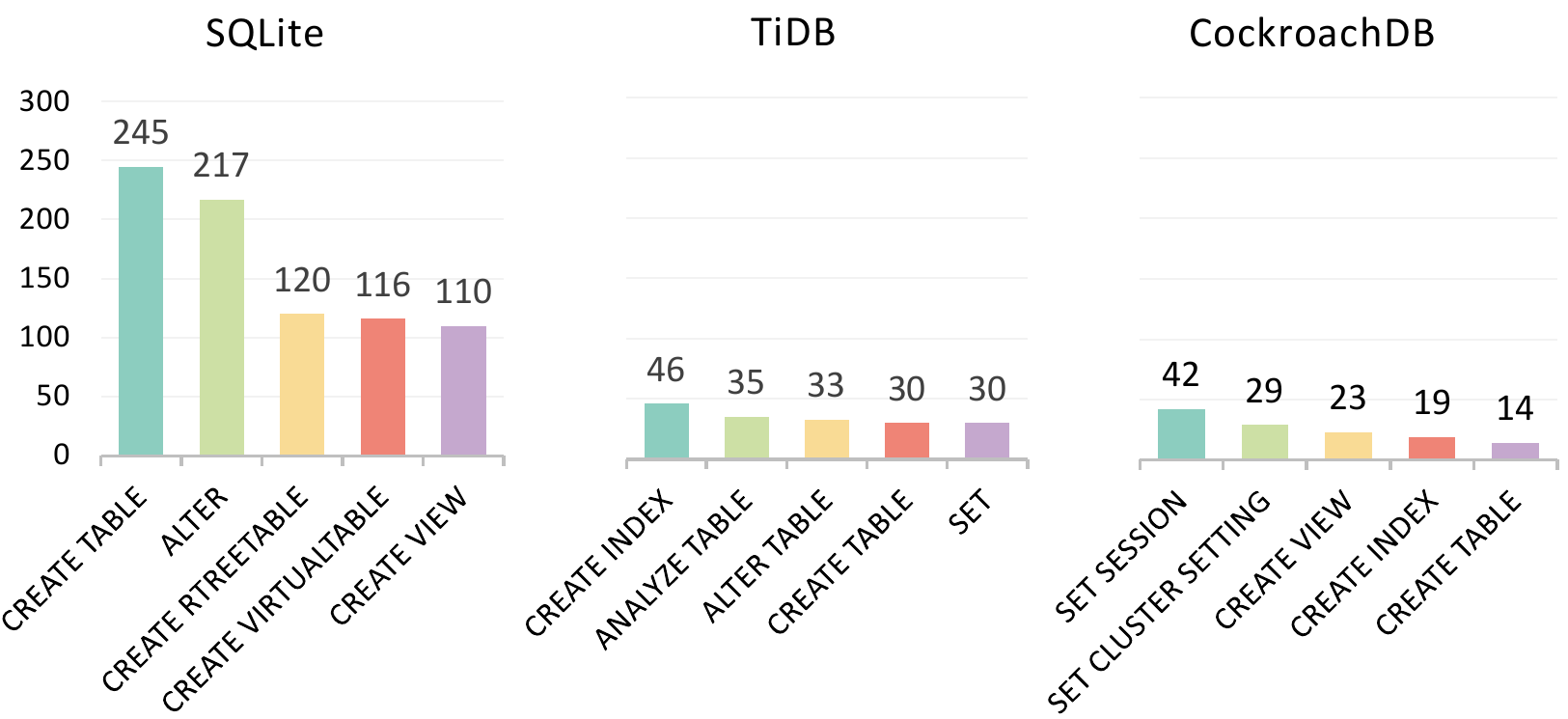}
    \vspace{-6mm}
    \caption{How often a mutation was executed for the five most frequently executed mutations for SQLite, TiDB, and CockroachDB across 10 runs.}
    \label{fig:mutator}
\end{figure}

\emph{Sensitivity of mutations.}
 To evaluate the contribution of each mutation, we examined how often each mutation (\emph{i.e.}, SQL statement) was executed across 10 runs in 24 hours. \Autoref{fig:mutator} shows the five most frequently executed mutations for each DDBMS. The most frequently-executed mutation for SQLite is \lstinline{CREATE TABLE}. Other frequently executed mutations either create other kinds of tables that are unique to SQLite or change the schema of existing tables using \lstinline{ALTER}. This is expected, as more kinds of tables subsequently cause SQLite to explore more query plans. Despite frequently creating additional tables, we did not observe excessive execution times, as we limited the maximum number of tables and indexes. For TiDB and CockroachDB, the number of mutations is much lower than that for SQLite, as we could run them for only up to 6 hours.
 \qpg favors the mutation \lstinline{CREATE INDEX} for TiDB, because indexes allow it to use more efficient physical operators when reading data. For CockroachDB, \qpg favors the mutation \lstinline{SET SESSION}, because it changes the system options, which can have an impact on the query plan. \qpg favors creating tables as various types of tables are supported in SQLite. Overall, all DBMSs have common frequent mutations, such as \lstinline{CREATE TABLE}, yet have distinct frequent mutations, such as \lstinline{SET}, depending on the various characteristics of DBMS.

For all three analyses, $\hat A_{12}=1$ and $U<0.05$ for \tool against \sqlancer on all DBMSs, which indicates the results are statistically significant.
\section{Related Work}

\emph{Fuzzing.} Fuzzing is an automatic software testing technique that generates or mutates inputs to target programs for finding crash bugs~\cite{miller1990empirical}. In recent years, it has gained increased attention, because of the success of the coverage-guided grey-box fuzzers such as AFL~\cite{afl, libfuzzer}, which instrument target programs to record code coverage which is subsequently used to mutate inputs to maximize code coverage. A plethora of works~\cite{recentfuzzing} have been proposed to improve fuzzing in various aspects. 
While QPG relates to grey-box fuzzing, we focus on finding logic bugs and DBMSs specifically, and guide test case generation by query plans rather than code coverage.

\emph{Finding logic bugs.} Various techniques have been proposed to find logic bugs in DBMSs. Differential testing~\cite{mckeeman1998differential} is a general technique that compares the outputs of multiple systems for the same input to detect potential discrepancies indicating bugs; various approaches use it as a test oracle for finding logic bugs by using different DBMSs\cite{slutz1998rags, ghit2020sparkfuzz, sqllogictest} or different versions of a DBMS\cite{yan2018snowtrail, yagoub2008oracle}. While such approaches have successfully found bugs, they are prone to false alarms due to differences in SQL dialects or expected differences between versions. Subsequently, three test oracles were proposed and implemented in \sqlancer~\cite{Rigger2020PQS, Rigger2020TLP, Rigger2020NoREC}. 
While we evaluated our technique with the state-of-the-art oracles NoREC and TLP, our method is compatible with any oracles.

\emph{Query generation.} Targeted and random generations are two major directions in query generation. As for targeted query generation, Bati et al.~\cite{Bati2007} proposed to incorporate execution feedback, such as code coverage, for guiding query generation to reach a specific code location. Khalek et al.~\cite{abdul2010automated} used a solver-backed approach to generate syntactically and semantically correct queries. 
Generating queries that satisfy cardinality constraints has been proven to be computationally hard, which is why heuristic algorithms were proposed~\cite{bruno2006generating, mishra2008generating}. As for random query generation, SQLsmith~\cite{sqlsmith} uses a predefined grammar to randomly generate semantic valid queries and has found over 100 bugs in widely-used DBMSs. APOLLO~\cite{apollo} also uses a predefined grammar to generate queries for finding regression performance issues. Similarly, we use a grammar-based random generation method for generating valid queries. Squirrel~\cite{zhong2020squirrel} and \sqlright~\cite{liang:sqlright} use a mutation-based method to generate new queries, but such approaches are prone to generating queries that are semantically invalid.

\emph{Database state generation.} Similarly, targeted and random generations are two major directions in database state generation. As for targeted database state generation, QAGen~\cite{qagen} uses symbolic execution to specify constraints and generate queries that satisfy the constraints. SPQR~\cite{binnig2006reverse} generates the database state for a given query and expected results. As for random database state generation, Gray et al.~\cite{gray1994quickly} proposed to quickly generate billions-record databases using parallel algorithms. Coverage-based methods~\cite{zhong2020squirrel, liang:sqlright} generate new database states by mutating given SQL statements that are used to create the database state. Compared with these methods, we used query plans as guidance to generate more diverse database states for efficiently finding logic bugs.

\emph{Query plan in testing.} Database researchers have invested decades of effort to improve the performance of DBMSs, often by improving the performance of generated query plans or the operators used in them~\cite{graefe1993query, wu2013predicting, mchugh1999query, paul2016gpl, giceva2014deployment}; providing a comprehensive summary of these exceeds the scope of this paper. In terms of testing, Gu et al.~\cite{gu2012testing} proposed measuring the accuracy of query optimizers by forcing the generation of multiple alternative query plans for each test case, timing the execution of all alternatives, and ranking the plans by their effective costs with the goal of comparing this ranking with the ranking of the estimated cost.
Leis et al.~\cite{leis2015good} measured both the effects of the cost model and cardinality estimators used to derive an efficient query plan.
Rather than improving, studying, or testing the accuracy of query plans, we use query plans to guide test case generation.
\section{Conclusion}
In this paper, we have proposed the concept of \emph{Query Plan Guidance} (\qpg) to efficiently detect logic bugs in DBMSs. Its core insight is that the DBMS' internal execution logic for a given query is reflected by its query plan and, therefore, covering more unique query plans might increase the likelihood of finding logic bugs. Our study shows that the query plans of the queries in previously-found bugs vary significantly, but are simple. Thus, we designed an algorithm to gradually mutate database states toward more unique and complex query plans. \qpg enabled us to find \numbugs unique, previously unknown bugs in widely-used and extensively-tested database systems---SQLite, TiDB, and CockroachDB. The experiments show that \qpg results in \qpcovfactor more unique query plans than a random-generation method and \qpcovfactorsqlright more than a code coverage-guidance method. \qpg also improves logic-bug finding efficiency by 2$\times$.
Overall, this paper has demonstrated that \qpg is a general-applicable, black-box approach that increases bug-finding efficiency and enables finding difficult-to-trigger bugs.
While we demonstrated \qpg in the context of automated testing, we believe that the core idea could be applied also in other contexts (\emph{e.g.}, to measure the quality of a test suite).

\section{Data Availability}

Our implementation and experimental data are publicly available at  \url{https://zenodo.org/record/7553013}.

\bibliographystyle{IEEEtran}
\bibliography{references}

\end{document}